\documentclass[%
 aip,
 amsmath,amssymb,
 reprint,%
]{revtex4-1}

\usepackage{graphicx}
\usepackage{dcolumn}
\usepackage{bm}
\usepackage[utf8]{inputenc}
\usepackage[T1]{fontenc}
\usepackage{mathptmx}
\usepackage{etoolbox}
\usepackage{threeparttable}
\usepackage{array}
\usepackage{xspace}
\usepackage{multirow}
\usepackage{booktabs}
\usepackage[version=4]{mhchem}
\usepackage{xcolor}
\usepackage{natbib}

\newcolumntype{d}{D{.}{.}{5}}
\newcolumntype{e}{D{.}{.}{10}}
\newcommand{\wn}{cm$^{-1}$\xspace} 
\newcommand{\mrm}[1]{\ensuremath{\mathrm{#1}}}
\newcommand{\mcl}[3]{\multicolumn{#1}{#2}{#3}}
\newcommand{\mrw}[3]{\multirow{#1}{#2}{#3}}

\makeatletter
\def\@email#1#2{%
 \endgroup
 \patchcmd{\titleblock@produce}
  {\frontmatter@RRAPformat}
  {\frontmatter@RRAPformat{\produce@RRAP{*#1\href{mailto:#2}{#2}}}\frontmatter@RRAPformat}
  {}{}
}%
\makeatother
\begin{document}

\preprint{AIP/123-QED}

\title[\ce{HCO+} and \ce{He} collisional system]{
An improved study of
\ce{HCO+} and \ce{He}  system: interaction potential, collisional relaxation and pressure broadening}


\author{F. Tonolo}
 \affiliation{Scuola Normale Superiore, Piazza dei Cavalieri 7, I-56126 Pisa, Italy.}
 \affiliation{Dipartimento di Chimica “Giacomo Ciamician”, Università di Bologna, Via F. Selmi 2, I-40126 Bologna, Italy}

\author{L. Bizzocchi*}%
 \affiliation{Scuola Normale Superiore, Piazza dei Cavalieri 7, I-56126 Pisa, Italy.}
 \affiliation{Dipartimento di Chimica “Giacomo Ciamician”, Università di Bologna, Via F. Selmi 2, I-40126 Bologna, Italy}
 \email[Corresponding author: ]{luca.bizzocchi@unibo.it}

\author{M. Melosso}%
 \affiliation{Dipartimento di Chimica “Giacomo Ciamician”, Università di Bologna, Via F. Selmi 2, I-40126 Bologna, Italy}

\author{F. Lique*}
 \affiliation{Univ. Rennes, CNRS, IPR (Institut de Physique de Rennes) – UMR 6251, F-35000 Rennes, France}
  \email[Corresponding author: ]{francois.lique@univ-rennes1.fr} 

\author{L. Dore}
 \affiliation{Dipartimento di Chimica “Giacomo Ciamician”, Università di Bologna, Via F. Selmi 2, I-40126 Bologna, Italy}
 
\author{V. Barone}
 \affiliation{Scuola Normale Superiore, Piazza dei Cavalieri 7, I-56126 Pisa, Italy.} 

\author{C. Puzzarini* }
 \affiliation{Dipartimento di Chimica “Giacomo Ciamician”, Università di Bologna, Via F. Selmi 2, I-40126 Bologna, Italy}
  \email[Corresponding author: ]{cristina.puzzarini@unibo.it}

 \homepage{http://www.Second.institution.edu/~Charlie.Author.}

\date{\today}

\begin{abstract}
In light of its ubiquitous presence in the interstellar gas, the chemistry and reactivity of the \ce{HCO+} ion requires special attention. 
The availability of up-to-date collisional data between this ion and the most abundant perturbing species in the interstellar medium is a critical resource in order to derive reliable values of its molecular abundance from astronomical observations.
This work intends to provide improved scattering parameters for the \ce{HCO+} and \ce{He} collisional system.
We have tested the accuracy of explicitly correlated coupled–cluster methods for mapping the short– and long–range multi–dimensional potential energy surface of an atom--ion systems.
A validation of the methodology employed for the calculation of the potential well has been obtained from the comparison with experimentally derived bound-state spectroscopic parameters. 
Finally, by solving the close-coupling scattering equations, we have derived the pressure broadening and shift coefficients for the first six rotational transitions of \ce{HCO+}, as well as inelastic state-to-state transition rates up to $j=5$ in the 5--100\,K temperature interval.
\end{abstract}

\maketitle

\section*{Introduction}
The harsh conditions of the interstellar medium (ISM) pose severe constraints to the chemical processes it hosts, which exhibit behaviors that greatly differ from those occurring in terrestrial environments.
For instance, in space, the molecular energy level populations are rarely at local thermodynamic equilibrium (LTE) since the density is usually so low ($\sim 10^2 - 10^6$ cm$^{-3}$) that collisions compete with radiative processes.
Under such conditions, the estimate of molecular abundances in the ISM from spectral lines requires the knowledge of their collisional coefficients for the most abundant perturbing species. 
Their nature depends on the investigated intestellar environment, being for most cases neutral species like \ce{H2} or \ce{He} \cite{roueff2013molecular}, but also collisions with electrons should be considered in photon dominated regions \cite{kauffmann2017molecular}. 
In this context, the study and computation of the collisional parameters has gained an increasing interest, the aim being the balance between accuracy and computational cost.

In this work, we benchmarked the performance of different levels of theory in describing the interaction potential of a collisional system. 
With the aim of extending the discussion also to larger systems, we kept a keen eye on the computational cost. 
A remarkable outcome in this regard exploits the good perfomances of explicitly correlated coupled–cluster methods \cite{adler2007simple,knizia2009simplified,peterson2008systematically,kendall1992electron,lique2010benchmarks,ajili2013accuracy} for mapping the short– and long–range multi–dimensional potential energy surface (PES) of collisional systems.

The system we opted to investigate addresses the collision between the \ce{HCO+} ion with \ce{He}.
As far as we know, this study reports the first application of explicitly correlated methods to this collisional system and represents the most accurate description of the underlying interaction potential - whilst maintaining an affordable computational cost. 
Moreover, given the great relevance of ion chemistry in terms of the molecular evolution of the interstellar medium, the \ce{HCO+} ion is particularly interesting \cite{petrie2007ions}: it is the most abundant cation in dense molecular clouds \cite{herbst2005molecular} and has been detected in a large number of objects with widely differing physical characteristics \cite{snyder1976detection,langer1978observations,welch1981millimeter} (see also Ref.~\citenum{lattanzi2007rotational} for an exhaustive list of recent detections). For this reason, it has a prominent role when seeking for new interstellar chemical networks and has been often used as a tracer of ionization in different dense interstellar cores \cite{caselli1998ionization}. 

The first interstellar detection of the \ce{HCO+} ion dates back to 1970 \cite{buhl1970unidentified}, albeit its identification was not unambiguously verified until the characterization of its rotational spectrum in 1975 \cite{article}.
Given its astrochemical relevance, several previous studies have determined the experimental and computational counterparts for some of the \ce{HCO+} scattering parameters. 
The first set of rotational de-excitation rate coefficients of \ce{HCO+} in collision with both $para-$ and $orto-$H$_2$ were recently determined by Alpizar $et$ $al.$ \cite{denis2020rotational}
As regards the \ce{HCO+} and \ce{He} collisional system, the computation of the first state-to-state rate coefficients dates back to 1985 \cite{monteiro1985rotational}. Afterwards, in 2008, Buffa $et$ $al.$ \cite{buffa2008experimental} characterized a new PES by employing the CCSD(T) method (coupled cluster within singles, doubles and a perturbative treatment of triples excitations) \cite{raghavachari1989fifth} in conjunction with a quadruple-$\zeta$ quality basis set (aug-cc-pVQZ) \cite{peterson1994benchmark}, from which the pressure broadening and pressure shift parameters were then derived \cite{buffa2008experimental}. In the same work the experimental results obtained by means of a frequency modulated spectrometer for three rotational lines of \ce{HCO+} at 88 K have been reported. Shortly after, starting from the same PES, the corresponding state-to-state rates have also been computed \cite{buffa2009state}.
Lastly, in 2019, Salomon et $al.$ experimentally determined, employing the double resonance technique, the rotational parameters associated with the ground state of the bound system \ce{He-HCO+} \cite{Salomon2019}.
The comparative analysis of such data with those obtained in this work has therefore permitted a robust validation of the employed methodologies.

This paper is organized as follow.
Section \ref{sec1} provides an in-depth analysis leading to the choice of the level of theory for the description of the PES of the collisional system. Section \ref{sec2} shows the subsequent derivation of the rotational parameters of bound state, which led to a further validation of the computed potential by comparison with the experimental results.
Finally, section \ref{sec3} describes the performed quantum scattering computations and the derived parameters: the computation of the inelastic cross section is detailed in section \ref{sec3a}, while the inelastic rate coefficients between the rotational states of the system and the 
pressure broadening and pressure shift coefficients are presented in sections \ref{sec3b} and \ref{sec3c}, respectively.

\section{Construction and testing of the potential energy surface \label{sec1}}

The starting point of our study on the \ce{HCO+} and \ce{He} collisional system is the accurate investigation of its intermolecular PES.
We describe the system using standard Jacobi coordinates, i.e., the distance between the center of mass of \ce{HCO+} and the He atom ($R$) and the angle $\theta$ between the molecular axis and $R$ distance vector.

\begin{figure}[h!]
 \centering
 \includegraphics[scale=0.17]{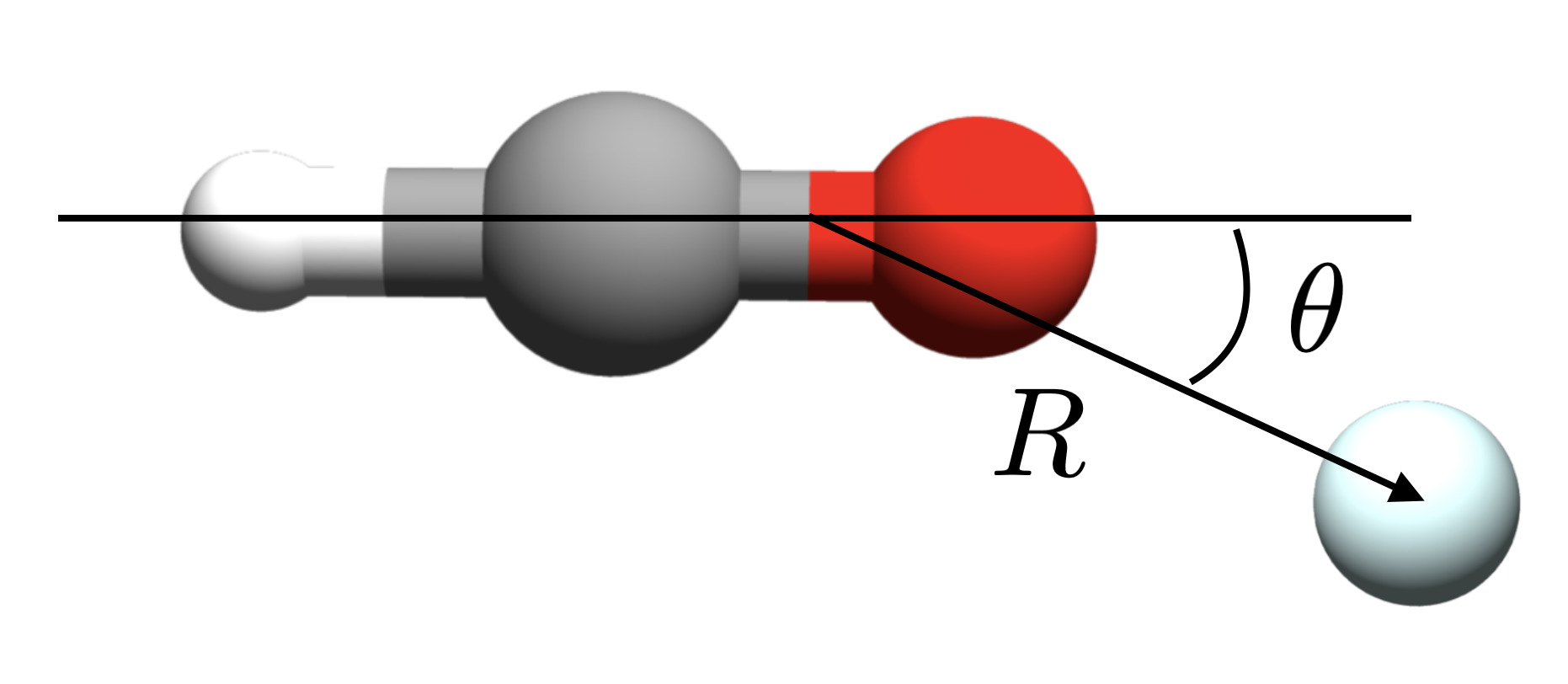}
 \caption{\footnotesize{Jacobi internal coordinates of the \ce{HCO+} and \ce{He} collisional system.}}
 \label{eulero}
\end{figure}

The lowest vibrational mode of the \ce{HCO+} ion lies at \mbox{$\sim 829$\,\wn} (see Ref.~ \citenum{davies1984diode}) so that, under the non-reactive low-temperature conditions which we target in the present investigation, all the vibrational channels can be safely considered as closed.
Accordingly, the \ce{HCO+} structure was held fixed to its experimentally determined $r_e$ geometry \cite{dore2003struct}, linear and with \ce{C-H} and \ce{C-O} bond distances of 1.0920 and~1.1056\,\r{A}, respectively.

\begin{table*}[ht]
 \renewcommand{\tabcolsep}{1pt}
 \renewcommand{\arraystretch}{1.1}
 \scriptsize
 \caption{CP-corrected interaction energies (\wn) for different geometries of the \ce{HCO+} and \ce{He} collisional system.}
 \begin{ruledtabular}
  \begin{tabular}{cc ddm{0.01cm}ddm{0.01cm}dm{0.01cm}dd}
   \mcl{2}{c}{\mrw{4}{*}{Geometry}} & 
   \mcl{7}{c}{CCSD(T)} & & \mcl{2}{c}{CCSD(T)-F12$^\text{a}$} \\
   \cmidrule{3-9} \cmidrule{11-12}
    & & 
    \mcl{2}{c}{Peterson CBS Extrapolation\footnote{Extrapolation to the CBS limit of the fc-CCSD(T) energies performed with the Peterson three-points extrapolation formula with $n = 3,4,5$.}} 
    & & 
    \mcl{2}{c}{Feller+Helgaker CBS Extrapolation\footnote{The extrapolation of the HF-SCF energy performed with the three-point formula by Feller ($n=3, 4, 5$), combined with the extrapolation of the fc-CCSD(T) correlation energy using the two-point ($n=3,4$) formula by Helgaker.}} 
    & & \mcl{1}{c}{\mrw{2}{*}{aug-cc-pVQZ}} 
    & & \mcl{1}{c}{\mrw{2}{*}{aug-cc-pVTZ}} 
    & \mcl{1}{c}{\mrw{2}{*}{jun-cc-pVTZ}} \\
    \cmidrule{3-4}\cmidrule{6-7}
    \mcl{1}{c}{R}           &  
    \mcl{1}{c}{$\theta$}    & 
    \mcl{1}{c}{aug-cc-pV$n$Z} & 
    \mcl{1}{c}{jun-cc-pV$n$Z} &  & 
    \mcl{1}{c}{aug-cc-pV$n$Z} & 
    \mcl{1}{c}{jun-cc-pV$n$Z} &  & & &  &  \\[1ex]
    \hline \\[-1ex]
    2.0  &   0.0  & 16670.07  & 16674.57  & & 17812.85  & 17689.38 &  & 16763.42   & & 16717.49  & 16752.53   \\
    3.5  &   0.0  & -66.41    & -61.07    & & -8.77     & -24.23   &  & -61.02     & & -69.90    & -60.37     \\                5.0  &   0.0  & -15.40    & -16.70    & & -8.41     & -8.83    &  & -15.66     & & -15.64    & -14.17     \\
    7.5  &   0.0  & -0.90     & -2.20     & & -0.73     & -0.68    &  & -2.91      & & -2.90     & -2.36      \\
   10.0  &   0.0  & 1.30      & -0.01     & & 0.80      & 0.82     &  & -0.93      & & -0.97     & -0.78      \\
    2.0  &  45.0  & 6313.38   & 6315.28   & & 7348.92   & 7220.16  &  & 6370.79    & & 6340.73   & 6379.80    \\                3.5  &  45.0  & -70.39    & -66.35    & & -26.91    & -39.02   &  & -69.54     & & -71.53    & -61.78     \\                  
    5.0  &  45.0  & -14.09    & -16.70    & & -9.72     & -10.01   &  & -16.13     & & -16.03    & -14.10     \\    
    7.5  &  45.0  & -0.90     & -2.20     & & -0.96     & -0.84    &  & -3.08      & & -3.11     & -2.52      \\  
   10.0  &  45.0  & 1.30      & -0.01     & & 0.74      & 0.70     &  & -0.99      & & -1.04     & -0.83      \\
    2.0  &  90.0  & 2139.73   & 2139.50   & & 2741.54   & 2634.35  &  & 2177.23    & & 2152.64   & 2196.11    \\
    3.5  &  90.0  & -91.50    & -85.25    & & -59.17    & -66.54   &  & -89.72     & & -91.22    & -81.39     \\
    5.0  &  90.0  & -18.49    & -18.90    & & -15.30    & -14.50   &  & -20.46     & & -20.26    & -18.10     \\
    7.5  &  90.0  & -4.40     & -5.70     & & -1.87     & -1.52    &  & -3.84      & & -3.82     & -3.19      \\
   10.0  &  90.0  & 1.30      & -2.20     & & 0.57      & 0.52     &  & -1.19      & & -1.23     & -1.00      \\
    2.0  & 135.0  & 9549.84   & 9543.82   & & 10301.07  & 10195.37 &  & 9623.74    & & 9578.38   & 9620.42    \\
    3.5  & 135.0  & -178.60   & -171.79   & & -123.26   & -138.63  &  & -171.77    & & -174.03   & -158.42    \\
    5.0  & 135.0  & -33.90    & -32.09    & & -29.97    & -28.07   &  & -35.98     & & -35.93    & -32.66     \\
    7.5  & 135.0  & -3.10     & -4.40     & & -3.60     & -3.02    &  & -5.52      & & -5.48     & -4.72      \\
   10.0  & 135.0  & -2.20     & -2.20     & & 0.16      & 0.25     &  & -1.55      & & -1.58     & -1.31      \\
    2.0  & 180.0  & 191024.32 & 191019.68 & & 193450.01 & 193365.57&  & 191419.52  & & 191286.67 & 191318.90  \\
    3.5  & 180.0  & -269.65   & -271.98   & & -126.56   & -155.74  &  & -265.64    & & -268.54   & -250.38    \\
    5.0  & 180.0  & -60.70    & -50.55    & & -49.50    & -48.73   &  & -58.24     & & -58.93    & -53.99     \\
    7.5  & 180.0  & -5.29     & -7.90     & & -5.01     & -4.28    &  & -6.97      & & -6.84     & -6.03      \\
   10.0  & 180.0  & -2.20     & -2.20     & & -0.17     & 0.00     &  & -1.80      & & -1.80     & -1.53      \\ 
\hline    \\[-1.5ex]
\mcl{2}{c}{CPU time \footnote{Mean CPU time (s) needed to compute one point of the energy grid (rounded values).}} &  9000 &  7400 & &  8500 & 4300 & &  8500 & &  2600 &  1700 \\
  \end{tabular}
\end{ruledtabular}

\label{bench}
\end{table*}

 \begin{figure*}[t]
 \centering
 \includegraphics[scale=0.69]{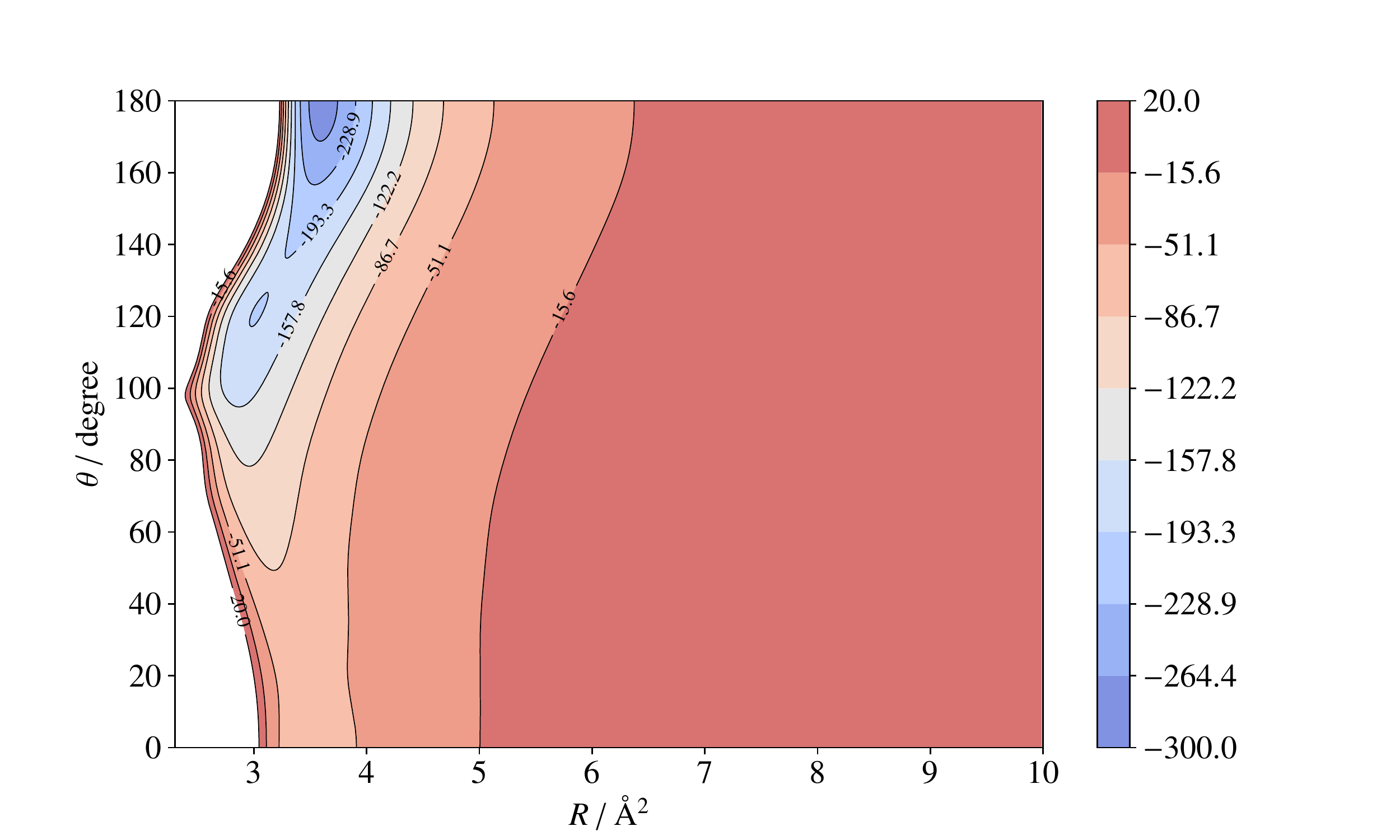}
 \caption{Contour plot of the \ce{HCO+} and \ce{He} interaction PES as a function of $$R$$ and $\theta$. Energies are in cm$^{-1}$. \label{pes2d}}
\end{figure*}

For the ab initio calculation of the interaction energy, the choice of the level of theory that best combines accuracy and computational efficiency was guided by a preliminary benchmark study on a sample of 25 geometries that tested different methodologies and basis sets. 
The results are reported in Table \ref{bench}.

Given the strong ionic effect of \ce{HCO+}, a sufficiently flexible basis set is needed to describe the electronic behavior in regions far from the electronic density maximum. For this reason, two triple-$\zeta$ correlation consistent basis sets \cite{dunning1989gaussian}, which introduce diffuse functions in a complete (aug-) or partial (jun-) way, have been evaluated \cite{kendall1992electron,woon1993gaussian,papajak2011perspectives}. As has been shown in many molecular systems involving non-covalent interactions, the partial addition of diffuse functions accurately predicts the electronic behavior while saving computational cost \cite{alessandrini2019extension}.
The chosen basis sets are the aug-/jun-cc-pV$n$Z where the method employed for all computations is the CCSD(T)  \cite{watts1993coupled}. 
In addition, a bi-electronic distance dependence in Slater-type form can be included in the electronic wave function. This contribution enhances the performance of the wavefunction for small interelectronic separations and has been demonstrated in many cases to be very suitable for mapping the short- and long-range multi-dimensional PESs \cite{lique2010benchmarks,ajili2013accuracy}. 
The methods that introduce this contribution are the so-called explicitly correlated methods, with the F12 approximation being employed. \cite{adler2007simple,knizia2009simplified,peterson2008systematically}.
In detail, the CCSD(T)-F12a method has been used in combination with both jun-/aug-cc-pVTZ basis sets, and within the frozen core approximation (fc).
All calculations were carried out with the MOLPRO program suite \cite{werner2012wires}.

Going into the details of the test: the CCSD(T)-F12 energies were compared with the ones obtained via the CCSD(T)/CBS composite scheme in which extrapolation to the complete basis set (CBS) limit has been achieved according to two different procedures, having different computational costs. 
In the first approach, the total energy is defined as: 
\begin{equation}\label{composito}
  E_\text{tot}=E^{\infty}_\text{HF} + \Delta E^{\infty}_\text{CCSD(T)} \,.
\end{equation}
Here, the first term on the right-hand side is the HF-SCF energy, extrapolated to the CBS limit by means of the Feller's exponential formula\cite{feller1992application}:
\begin{equation}\label{feller}
 E^{\infty}_\text{HF} = E^{n}_\text{HF} - B\,\mrm{e}^{-Cn}, \quad n=3,4,5 \,.
\end{equation}

The second term of Eq.~\eqref{composito} accounts for the extrapolation to the CBS limit of the CCSD(T) correlation energy ($E_\text{corr}$), within the fc approximation, using the 2-point \emph{n}$^{-3}$ formula by Helgaker and co-workers \cite{helgaker1997basis}:
\begin{equation}\label{helgaker}
 \begin{aligned}
  &\Delta E^{\infty}_\text{CCSD(T)}= \frac{n^{3} E_\text{corr}^n - (n-1)^3 E_\text{corr}^{n-1}}{n^3-(n-1)^3} \,,
 \end{aligned} 
\end{equation}
where $n=4$.

Alternatively, the total energy is obtained by applying the mixed exponential and Gaussian formula by Peterson \textit{et al.} \cite{peterson1994benchmark}:
\begin{equation}
 \begin{aligned}
 &E_{n}=E_{\text{CBS}}+\alpha\mrm{e}^{-(n-1)}+\beta\mrm{e}^{-(n-1)^2} \,, 
 \end{aligned}  
\end{equation}
where E$_{\mrm{CBS}}$, $\alpha$ and $\beta$ are adjustable parameters, and $n = 3,4,5$. The energies derived via this extrapolation on fully augmented basis sets were taken as the reference energies for the benchmark test.

The interaction energy $E_{\textrm{int}}$ has been determined as follows: 
\begin{equation}
  E_{\text{int}} = E_\mrm{AB} - (E_\mrm{A} + E_\mrm{B}) \,,
\end{equation} 
where $E_\mrm{AB}$ is the molecular complex energy, while $E_\mrm{A}$ and $E_\mrm{B}$ are the energies of the two fragments.
The interaction energies have also been corrected by a counterpoise (CP) contribution in order to balance out the energy overestimation given by the basis set superposition error (BSSE). The CP correction is computed using the Boys $\&$ Bernardi formula \cite{boys1970calculation}:
\begin{equation}
 \Delta E_{\text{CP}}= (E^{\mrm{AB}}_{\mrm{A}} - E^{\mrm{A}}_{\mrm{A}}) + (E^{\mrm{AB}}_{\mrm{B}} - E^{\mrm{B}}_{\mrm{B}}),
\end{equation}
where $E^{\mrm{AB}}_{\mrm{X}}$ is the energy of the monomer calculated with the same basis functions used for the cluster and $E^{\mrm{X}}_{\mrm{X}}$ is the energy of the monomer computed with its own basis set (X = A,B).

Inspection of Table~\ref{bench} illustrates the remarkably good performances of the F12-explicitly correlated methods, which provide a description of long- and short-range energy interactions in good agreement with that obtained via the computationally expensive Peterson CBS extrapolation scheme ($n=3,4,5$). Going into detail, the short-range energy comparison reveals a mean percentage error around 5\%. At long range, the low value of the energies make the percentage error comparatively higher. However, the difference between the energies is always lower than $2.5$~\wn .
On the other hand, the cheaper Feller \& Helgaker composite scheme, which exploits smaller basis sets ($n=3,4$) for the extrapolation of the CCSD(T) correlation energy,  fails to predict the energy trend with the same accuracy. 
Table~\ref{bench} also shows that the F12 method in combination with a triple-$\zeta$ quality basis set, performs slightly better than the conventional CCSD(T) model in conjuction with the aug-cc-pVQZ basis set. To date, the latter level of theory provides the most accurate PES available in the literature for the collisional system of interest \cite{buffa2008experimental}.

A further remarkable feature is that the use of partially augmented (-jun) basis does not significantly affect the description of the PES except, as expected, in the long-range regions, where the dispersive interactions provide a major contribution to the energy. Given the lower computational cost entailed, however, the jun-cc-pVTZ basis set may be still recommended for systems whose long-range interactions are less prominent.

On the basis of this benchmark test, in the present work the PES of the \ce{HCO+} and \ce{He} collisional system has been entirely investigated by employing the CCSD(T)-F12a/aug-cc-pVTZ model, which is the level of theory that offers the best compromise between accuracy and computational cost.
The interaction potential has been built from an irregular grid in the {$R$, $\theta$} coordinates. A total of 390 points has been chosen by sampling the portion of the PES for $R$ varying between 2 and 10\,\r{A} and for 13 $\theta$ values equally spaced throughout the molecular plane. The radial mesh is denser in the region between 2 and 4\,\r{A} in order to sample the energy behavior in the proximity of the potential well, where sizable anisotropic effects are expected.
Since we are dealing with a new potential energy surface for the system under investigation, a \texttt{.tar} archive containing the computed interaction energies for each set of Jacobi coordinates ($R,\theta$) is included in the Supporting Information.
 
For the solution of the nuclear Schr\"{o}dinger equation by means of the close coupling equations, it is useful to express the interaction potential as an expansion of angular functions. For an interaction system formed by a linear rigid rotor and an atom, we can define the potential as:
\begin{equation} \label{pot}
 V\left(R, \theta \right)=\sum_{\lambda}  v_{\lambda} (R) P_{\lambda} \left(\cos \theta \right) \,,
\end{equation}
where $P_{\lambda} \left(\cos \theta \right)$ is a Legendre polynomial and $v_{\lambda} (R)$ are the radial coefficients \cite{lique2019gas}. The polynomial expansion has been performed on 13 points ($\lambda_{\textrm{max}}$ = 12), i.e., on the number of $\theta$ angles at which the PES is sampled. Different $\lambda$ terms ($\lambda>0$) govern the magnitudes of the inelastic rotational transitions, allowing for changes of the molecular angular momentum by $\Delta j = \pm \lambda$. 
Likewise in all the molecular ion-atom collisions, the long-range parts of the potential are characterized by a sizable contribution due to the induction interactions. This contribution scales with the interparticle distance as $R^{-4}$ and is proportional to the square of the charge of \ce{HCO+} and to the static electric dipole polarizability of helium.

To ensure a correct behavior of the PES expansion at large distances, the radial coefficients $v_\lambda(R)$ have been fitted to the functional form:
\begin{multline} \label{vexp}
 v_\lambda(R) = \mrm{e}^{-a_1^\lambda R}\left(a_2^\lambda + a_3^\lambda R + a_4^\lambda R^2 + a_5^\lambda R^3\right) \\
 -\frac{1}{2}\left[1 + \tanh{R/R_\text{ref}}\right]
 \left(\frac{C^{\lambda}_4}{R^4} + \frac{C^{\lambda}_6}{R^6} + \frac{C^{\lambda}_8}{R^8} + \frac{C^{\lambda}_{10}}{R^{10}}\right) \,,
\end{multline}

where the $C^{\lambda}_n$ symbols are used to label the coefficients of the $R^{-n}$ terms. The hyperbolic tangent factor provides a smooth transition between the short-range region ($0 < R < R_\text{ref}$) where computed PES points are available and the long-range extrapolated domain ($R > R_\text{ref}$).

The analytic potential was found to accurately reproduce the calculated energies. The difference between the $ab$ $initio$ points and the values obtained from Eq.~\eqref{vexp} and the fitted radial coefficients is less than 1\% across the entire grid.
A contour plot of the potential derived from the fit is shown in Figure \ref{pes2d}. The potential shows a global minimum when helium is collinear with the collider and interacts with the hydrogen of \ce{HCO+} at $R$ = 3.6 \r{A}. The resulting interaction energy in this point is 279.78~\wn.

\section{Bound states \label{sec2}}
A way for validating the new computed PES of the collisional system is provided by the calculation of the bound-state energies, whose experimental values are available \cite{Salomon2019}. For this purpose, the BOUND program has been employed \cite{hutson2019user}. The reduced mass of the collisional system is 3.5171996\,amu, while the rotational energies of the ion have been computed from its experimental rotational parameters: $B = 1.48750100$\,\wn, $D = 2.76304\times 10^{-6}$\,\wn, and $H = 2.58\times 10^{-12}$\,\wn (Ref.~\citenum{cazzoli2012hco+}). Rotational states of \ce{HCO+} with $j$ in the 0--19 range have been included in the calculation and the resulting coupled equations have been solved using a log-derivative propagator with $R$ varying from~2 to~6\,\r{A}, and in an energy range between 0 and $-300$\,\wn.

\begin{table}
 \renewcommand{\tabcolsep}{1pt}
 \footnotesize
 \caption{Rotational transitions (MHz) of the \ce{He-HCO+} Van der Waals complex for $\nu_1 = 0$.}
 \label{bound}
 \centering 
 \begin{ruledtabular}
 \begin{tabular}{@{\hspace{4pt}}rr@{\hspace{12pt}} ddc}
  \mcl{2}{c}{$J''\quad\leftarrow\quad J'$} &
  \mcl{1}{c}{Computed\footnote{obtained by bound state calculations.}}   &  
  \mcl{1}{c}{Experimental\footnote{From Ref.~\citenum{Salomon2019}. Constants of higher order than $L$ are not reported.}}   & 
  \mcl{1}{c}{\% Error}             \\[0.5ex]
  \hline \\[-2ex]
     1  &  0   &  17381.45707	&  17395.1112       &   0.08  \\
     2  &  1   &  34755.08955	&  34782.5930       &   0.08  \\
     3  &  2   &  52104.04912	&  52154.8930       &   0.10  \\
     4  &  3   &  69454.38773	&  69504.5481       &   0.07  \\
     5  &  4   &  86754.36121	&  86824.3164       &   0.08  \\
     6  &  5   &  104023.1563	&  104107.0922      &   0.08  \\
     7  &  6   &  121235.5004	&  121345.9633      &   0.09  \\
     8  &  7   &  138402.6059	&  138534.1142      &   0.09  \\
     9  &  8   &  155532.4171	&  155664.8274      &   0.09  \\
    10  &  9   &  172563.4768	&  172731.3559      &   0.10  \\
    11  & 10   &  189540.7237	&  189726.8475      &   0.10  \\
    12  & 11   &  206444.7613	&  206644.3117      &   0.10  \\
    13  & 12   &  223231.2502	&  223476.4892      &   0.11  \\
    14  & 13   &  239966.9244	&  240215.7559      &   0.10  \\
    15  & 14   &  256559.9273	&  256853.9714      &   0.11  \\
    16  & 15   &  273065.8015	&  273382.4879      &   0.12  \\  
  \hline \\[-2ex]
  \mcl{2}{l}{\hspace{6pt}$B$}             &  8691.18(59) &  8698.1947(16)  &  0.08 \\
  \mcl{2}{l}{\hspace{6pt}$D$}             &  0.3221(79) &  0.318741(46)   & 1.07 \\
  \mcl{2}{l}{\hspace{6pt}$H\times 10^5$}  &   6.7(41)    &  10.03(6)     & 33.00  \\
  \mcl{2}{l}{\hspace{6pt}$L\times 10^7$}  &   -0.92(70)    &  -2.681(39)     & 65.67  \\
 \end{tabular}
 \end{ruledtabular}
 Note: numbers in parentheses are $1\sigma$ errors in units of the last quoted digit.
\end{table}

Table~\ref{bound} gathers the computed energy differences between the low-lying bound states with negative parity. These values correspond to the rotational transitions of the \ce{He-HCO+} van der Waals complex measured by Salomon \textit{et al.} \cite{Salomon2019} by double resonance spectroscopy in the ion trap apparatus, which are also reported in Tables~\ref{bound}.
The computed transitions have been fitted by expressing the level energies using the linear rotor expressions, as
\begin{multline} \label{linrot}
 E_J = BJ(J + 1) - DJ^2(J + 1)^2  \\
 + HJ^3(J + 1)^3 + LJ^4(J + 1)^4 \,,
\end{multline}

with $B$ being the rotational constants of the complex,  
and $D$ the quartic, $H$ the sextic and $L$ the octic centrifugal distortion coefficients.
Experimental and theoretical results compare quite well: the average percentage error between individual rotational frequencies is $\sim 0.1$\%, and the centrifugal trend of the rotational energy is also well reproduced, as demonstrated by the fair agreement for the quartic centrifugal distortion constants ($\sim 1$\%), and also for the very small sextic ($\sim 30$\%, i.e. they agree withing $5\sigma$).



\section{Scattering calculations \label{sec3}}
\subsection{Inelastic cross sections \label{sec3a}}


Having validated the well depth, the focus of this paper is to provide a new evaluation of the scattering quantities of the \ce{HCO+} and \ce{He} system. 
We thus solved the standard time-independent coupled scattering equations using the MOLSCAT program \cite{hutson2019user}.
Calculations were carried out at values of the kinetic energy ranging from 2 to 500\,\wn, with narrow steps at low energies (0.2\,\wn up to 50\,\wn and 0.5\,\wn up to 170\,\wn), gradually increasing to 5\,\wn up to 500\,\wn.
The propagation started at a minimum distance around 2\,\r{A} (i.e. where the repulsion barrier of the collisional system is located), whereas the long range limits have been chosen to ensure convergence of the inelastic cross sections over a given energy range.
The adopted type of propagator is the hybrid LDMD/AIRY \cite{alexander1984hybrid,alexander1987stable}. This hybrid approach combines the Manolopoulos diabatic modified log-derivative (LDMD) propagator \cite{manolopoulos1986improved}, operating at short range where the relevant variation in the potential requires strict steps for the propagation, and the Alexander-Manolopoulos Airy (AIRY) propagator \cite{alexander1987stable} at long range, which accounts for looser propagation steps. Such choice provides the best compromise between accuracy and computational efficiency. 
The rotational basis set has been adjusted in selected energy ranges to ensure convergence of
the inelastic cross sections. At the highest total energy considered in the present calculation (500\,\wn), the rotational basis was extended to $j = 32$. The maximum value of the total angular momentum $J = j + l$ used in the calculations was chosen to allow for the convergence of the inelastic cross sections within 0.005\,\r{A}$^2$.

\begin{figure}
 \begin{center}
  \includegraphics[scale=0.42]{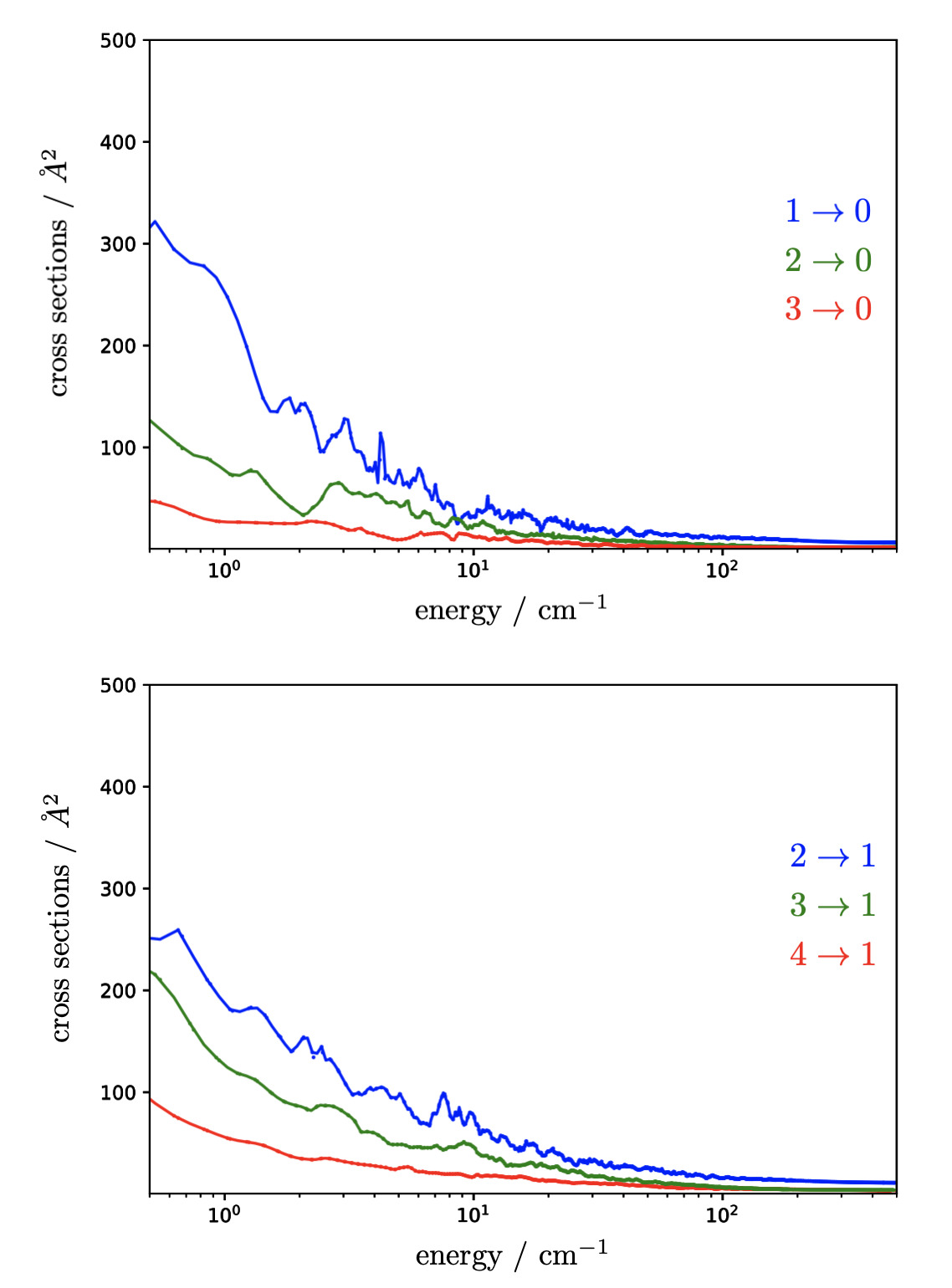}
 \end{center}
 \caption{Trends of some rotational de-excitation cross sections with collisional energy.}
 \label{ics}
\end{figure}

Figure \ref{ics} illustrates the energy dependence of the collisional de-excitation cross section for few selected rotational transitions. As expected, they show a general decrease as the collisional energy of the system increases. In Figure \ref{ics}, the portion up to 500 cm$^{-1}$ has been depicted, where oscillations due to different resonances are discernible for all the selected transitions. 
This trend is the same as that identified by Yazidi \textit{et al.} \cite{yazidi2014revised} for the \ce{HCO+} and \ce{H2} collisional system and is due to the presence of a potential well that supports many bound states.

\subsection{State-to-state transition rates \label{sec3b}}
\begin{figure*}
\begin{center}
\includegraphics[scale=0.31]{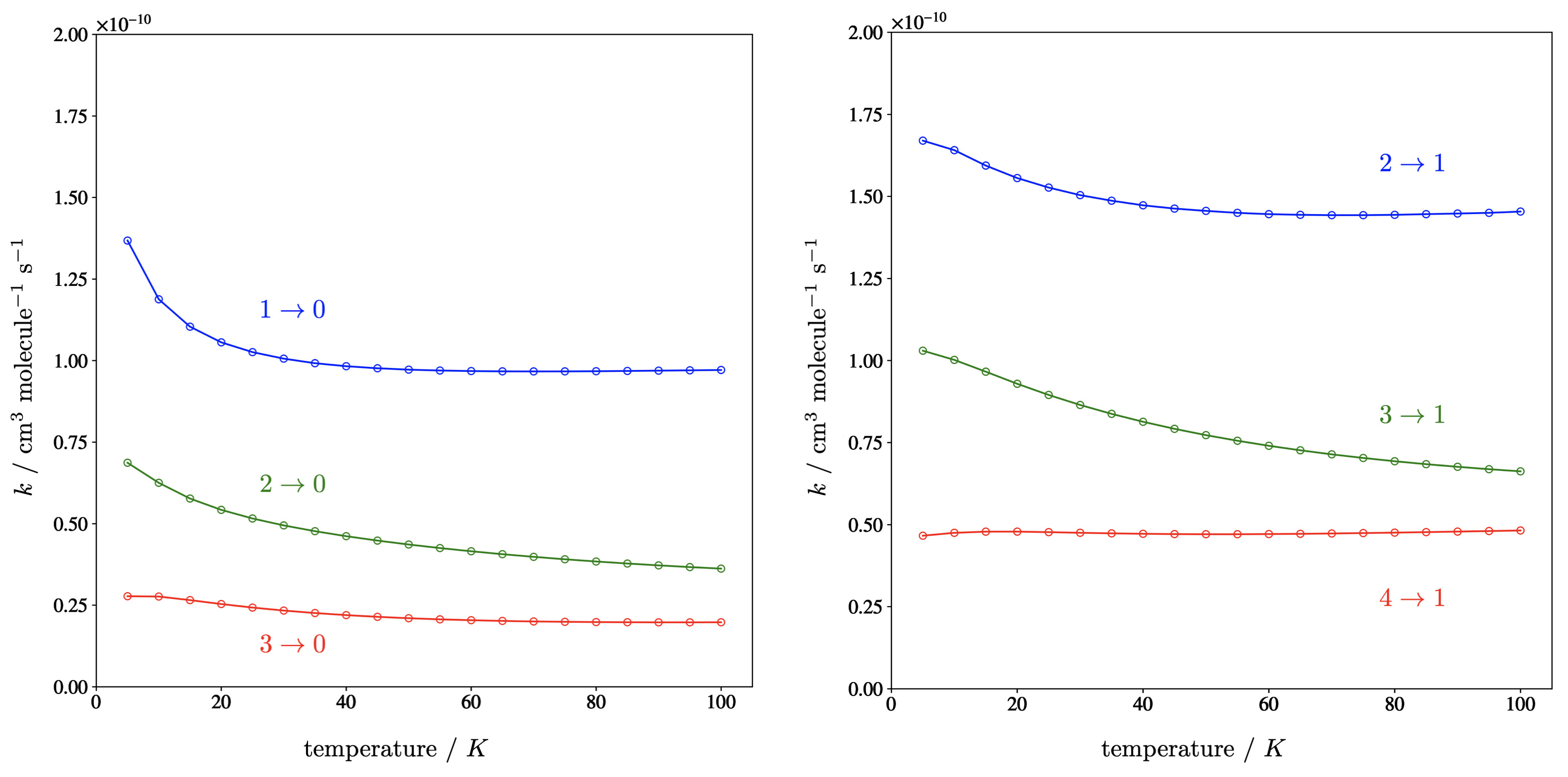}
\end{center}
\caption{Variation with temperature of some rotational de-excitation rate coefficients.}\label{rT}
\end{figure*}

The collisional calculations provide a set of inelastic cross sections as a function of the collision energy $\sigma(E_c)$. Starting from these quantities,  we have obtained the corresponding excitation and de-excitation rate coefficients, $k_{j^{\prime} \rightarrow j^{\prime\prime}}(T)$, for temperatures ranging from~5 to~100\,K. This was accomplished by averaging the $\sigma(E_c)$ over the collision energy:  
\begin{equation}
\begin{aligned}
 k_{j^{\prime} \rightarrow j^{\prime\prime}}(T) 
   &=\left(\frac{8}{\pi \mu k^{3} T^{3}}\right)^{1 / 2} \\
   &\times \int_{0}^{\infty} \sigma_{j^{\prime} \rightarrow j^{\prime\prime}}\left(E_{c}\right) E_{c} \exp \left(-E_{c} / k T\right) \mathrm{d} E_{c} \,,
\end{aligned}
\end{equation}
where $k$ is the Boltzmann constants and $\mu$ is the reduced mass of the system.
Some excitation rate coefficients are listed in Table \ref{rates}, thus enabling a comparison with those calculated in two previous works \cite{buffa2009state, monteiro1985rotational}. The agreement is good, especially when compared with the most recent results \cite{buffa2009state}, with an absolute maximum deviation of only $0.1 \times 10^{-10}$ cm$^3$ s$^{-1}$. 
\begin{table}[h!]
\renewcommand{\tabcolsep}{2pt}
\renewcommand{\arraystretch}{1.1}
    \scriptsize
    \caption{Transition rates for the excitation from $j^{\prime}$ to $j^{\prime\prime}$ at 10 K. Units are $10^{-10}$ cm$^3$ s$^{-1}$.}
    \centering 
    \begin{ruledtabular} 
    \begin{tabular}{lddd}
        
        \multicolumn{1}{c}{$j^{\prime} \rightarrow j^{\prime\prime}$} & \multicolumn{1}{c}{This work} & \multicolumn{1}{c}{Ref. ~\onlinecite{buffa2009state}} & \multicolumn{1}{c}{Ref.~\onlinecite{monteiro1985rotational}} \\
        \hline
        0  $\rightarrow$ 1    &  2.322	& 2.200  &  1.984   \\                                    
        0  $\rightarrow$ 2    &  0.866	& 0.857  &  0.739   \\                                    
        0  $\rightarrow$ 3    &  0.149	& 0.143  &  0.137   \\                                    
        0  $\rightarrow$ 4    &  0.022	& 0.021  &  0.014   \\                                    
        1  $\rightarrow$ 2    &  1.162	& 1.152  &  1.099   \\                                    
        1  $\rightarrow$ 3    &  0.275	& 0.266  &  0.195   \\                                    
        1  $\rightarrow$ 4    &  0.030	& 0.032  &  0.025   \\                                    
        2  $\rightarrow$ 3    &  0.643	& 0.657  &  0.566   \\                                    
        2  $\rightarrow$ 4    &  0.104	& 0.097  &  0.084   \\                                    
        3  $\rightarrow$ 4    &  0.390	& 0.405  &  0.310   \\                                    
        
    \end{tabular}
    \end{ruledtabular}
    \label{rates}
    \end{table}

Figure~\ref{rT} illustrates the temperature dependence of some de-excitation coefficients up to 100\,K, and the rate coefficients for the transitions $(1,2,3) \rightarrow 0$ and $(2,3,4) \rightarrow 1$. One can clearly see that, at low temperatures, the rate coefficients tend to decrease by increasing the energy. This trend significantly fades as $\Delta j$ increases, becoming hardly discernible for $\Delta j=3$. At higher temperature, all the rate coefficients become almost independent of the temperature, thus reflecting the prediction of the Langevin theory for ion-neutral interactions.
A comprehensive list of the integrated de-excitation rate coefficients up to $j=5$ is presented in 
Tables I and II the Supporting Information.

\subsection{Pressure broadening and shift \label{sec3c}}
\begin{figure*}
 \begin{center}
 \includegraphics[scale=0.32]{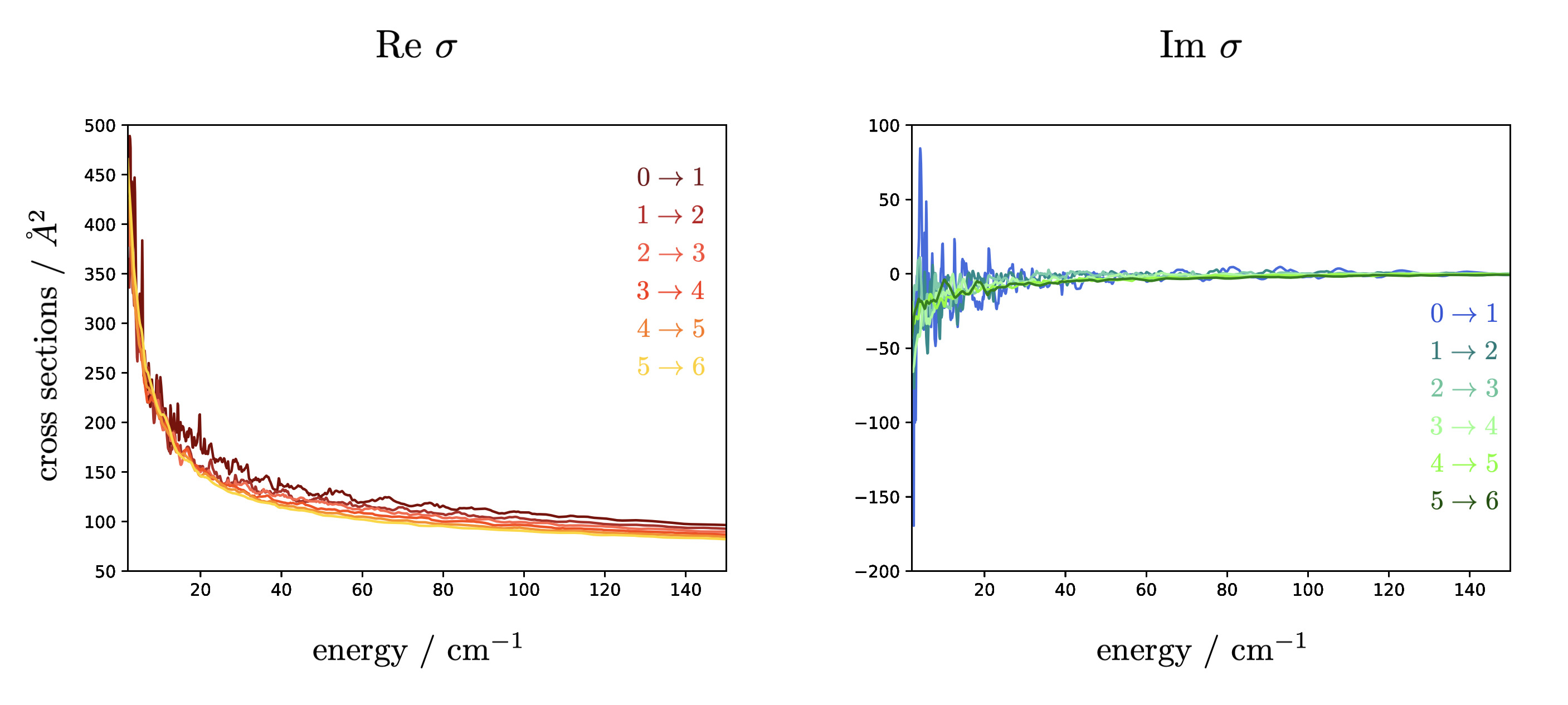}
 \end{center}
 \caption{Dependence of $\sigma$ on the thermal energy of the system for the six lowest rotational transitions.}
 \label{cross}
\end{figure*}

The pressure broadening and pressure shift coefficients were evaluated for the six lowest-energy rotational transitions, for which Buffa \textit{et al.} computed results and some experimental counterparts \cite{buffa2008experimental} are also available.

For this purpose, the $S$-matrices obtained from scattering calculations have been employed. The computation of these cross sections requires $S$-matrix elements involving both the initial and final states of the examined transition, which have the same kinetic energy but different total energies. 
The real (Re) and imaginary (Im) parts of the cross sections for a pair of upper and lower states describe the pressure broadening and shift, respectively, of a given $j^{\prime\prime} \leftarrow j^{\prime}$ line. 

\begin{table}[h]
\renewcommand{\tabcolsep}{1pt}
\renewcommand{\arraystretch}{1.1}
    \scriptsize
    \caption{Computed broadening cross section obtained by integration over the thermal energy distribution for the \ce{HCO+} and \ce{He} system at 88 K.}
    \vspace{0.5cm}
    \begin{ruledtabular} 
    \begin{tabular}{cdddd}
    \multicolumn{1}{c}{$j^{\prime\prime} \leftarrow j^{\prime}$} & \multicolumn{2}{c}{This work / [\AA$^2$]} & \multicolumn{2}{c}{{Ref.~\onlinecite{buffa2008experimental}} / [\AA$^2$]} \\
                                                                 & \multicolumn{1}{c}{Re $\sigma$} & \multicolumn{1}{c}{Im $\sigma$}  & \multicolumn{1}{c}{Re $\sigma$} & \multicolumn{1}{c}{Im $\sigma$} \\

    \hline
    1  $\leftarrow$ 0    & 113.120  & 1.354 &  110.05 & 1.23  \\                           
    2  $\leftarrow$ 1    & 106.850  & 1.205 &  104.06 & 1.33  \\             
    3  $\leftarrow$ 2    & 104.158  & 1.056 &  101.04 & 1.12  \\ 
    4  $\leftarrow$ 3    & 101.160  & 1.622 &  98.19  & 1.83  \\ 
    5  $\leftarrow$ 4    & 98.507   & 2.378 &  95.66  & 2.51  \\
    6  $\leftarrow$ 5    & 96.459   & 2.701 &  93.79  & 2.89  \\                                                               
\end{tabular}
\end{ruledtabular}
%
\label{cs}
\end{table}

\begin{table*}
\renewcommand{\tabcolsep}{1pt}
\renewcommand{\arraystretch}{1.1}
    \scriptsize
    \caption{Measured and calculated pressure broadening and shift parameters for the \ce{HCO+} and \ce{He} system at 88 K.}
    \begin{ruledtabular} 
    \begin{tabular}{cdcdddd}
    
    \multicolumn{1}{c}{$j^{\prime\prime} \leftarrow j^{\prime}$} & \multicolumn{1}{c}{Frequency} & \multicolumn{1}{c}{Parameter} & \multicolumn{1}{c}{Exp. Values\tnote{a}} & \multicolumn{1}{c}{This work} & \multicolumn{1}{c}{Recomputed values\footnote{Recomputed values from the cross sections taken from Table 2 of Ref.~\onlinecite{buffa2008experimental}.}} & \multicolumn{1}{c}{{Ref.~\onlinecite{buffa2008experimental}}}\\
                                                                 & \multicolumn{1}{c}{[MHz]} &                                   & \multicolumn{1}{c}{[MHz/Torr]} & \multicolumn{1}{c}{[MHz/Torr]} & \multicolumn{1}{c}{[MHz/Torr]} & \multicolumn{1}{c}{[MHz/Torr]}\\

    \hline
    1  $\leftarrow$ 0    &  89188.5261	&  broadening       &            & 14.377  & 13.987 & 13.76  \\ 
                         &              &  shift            &            & 0.172   & 0.156  & 0.154  \\                          
    2  $\leftarrow$ 1    &  178375.0642	&  broadening       &            & 13.580  & 13.225 & 13.01  \\ 
                         &              &  shift            &            & 0.153   & 0.169  & 0.168  \\            
    3  $\leftarrow$ 2    &  267557.6263	&  broadening       &            & 13.238  & 12.841 & 12.64  \\ 
                         &              &  shift            &            & 0.134   & 0.142  & 0.134  \\
    4  $\leftarrow$ 3    &  356734.2246	&  broadening       & 12.39(29)  & 12.857  & 12.479 & 12.27  \\ 
                         &              &  shift            & 0.328(19)  & 0.206  & 0.233  & 0.229  \\
    5  $\leftarrow$ 4    &  445902.8713	&  broadening       & 12.42(22)  & 12.520  & 12.158 & 11.95  \\
                         &              &  shift            & 0.427(29)  & 0.302  & 0.319  & 0.312  \\
    6  $\leftarrow$ 5    &  535061.5791	&  broadening       & 12.13(29)  & 12.260  & 11.920 & 11.72  \\
                         &              &  shift            & 0.497(17)  & 0.343  & 0.367  & 0.364  \\                                                                  
\end{tabular}

\end{ruledtabular}
%
\label{pb}
\end{table*}

The trend of the cross sections over the energy distribution shows irregular oscillations for all the studied transitions. This behavior is discernible from Figure~\ref{cross}, where the oscillations are clearly visible in the lower energy portion and are particularly pronounced for the transitions involving the low lying states. 
The final cross sections were calculated by integrating over the entire distribution of thermal energy:
\begin{equation} \label{integr}
 \bar{\sigma}=\frac{1}{(k T)^{2}} \int_{0}^{\infty} E \mathrm{e}^{-E/kT} \sigma(E) dE \,,    
\end{equation}
where $k$ is the Boltzmann constant and $T$ is the temperature chosen for the integration, which was set to 88\,K to allow the comparison with previously calculated and observed values.
The resulting pressure broadening cross sections are shown in Table~\ref{cs}, where the previous results by Buffa \textit{et al.} \cite{buffa2008experimental} are also reported.

The associated pressure broadening ($\Gamma$) and pressure shift ($s$) coefficients, are obtained from the real and imaginary part of the cross sections as:
\begin{equation} \label{pcoeff}
  \Gamma - \mathrm{i}s = n_{p} \bar{\nu} \bar{\sigma} =\frac{56.6915}{\sqrt{\mu T}} \bar{\sigma} \,,
\end{equation}
$\bar{\nu}=\left(8 k_{B} T / \pi \mu\right)^{1 / 2}$ is the mean velocity of the colliders and $n_p$ represents the density of the gas. In terms of units, $\Gamma$ and $s$ are expressed in cm$^{-3}$\,atm$^{-1}$, $\bar{\sigma}$ in A$^2$, $\mu$ (reduced mass of the system) in amu and $T$ in Kelvin. 

The resulting coefficients are listed in Table \ref{pb}, where the comparison with the experimental and previously computed counterparts is also reported.
A first noteworthy point is a discrepancy observed between the coefficients presented in Ref.~\citenum{buffa2008experimental} and the coefficients recalculated from the cross sections obtained by the same paper via Eq.~\eqref{pcoeff}. Both values have been reported in the last two columns of Table~\ref{pb}. This discrepancy may be attributable to an error in the cross section conversion since it has a systematic value on almost all transitions. Indeed, by reconverting the cross sections at 91\,K instead of 88\,K, our results became very close to those reported by Buffa \textit{et al.} \cite{buffa2008experimental} 

The comparison of the computed data with those experimentally measured revealed a remarkable agreement.
It appears that the PES computed in this work describes more accurately the pressure-broadening parameters related to the two higher energy transitions, where the percentage error with respect to the experiments is around $1 \%$. A somewhat larger discrepancy is exhibited by the $4 \leftarrow 3$ transition, for which the percentage error is $ \sim 4 \%$.
On the other hand, pressure shift coefficients, derived from the imaginary part of the broadening cross sections, show higher deviations compared to experiments. It should be noted, however, that such measurements are rather delicate and thus affected by significant uncertainties.
Nevertheless, from a qualitative point of view, their values show a good agreement with the ones computed in this work.




\section{Conclusions}
In this work, the evaluation and validation of an accurate computational procedure for the characterization of collisional potential energy surfaces has been reported. 
This investigation drew on the excellent performance of explicitly correlated methods for the description of short- and long-range interaction energies, whose affordable computational cost would also allow a straightforward extension of this methodology to larger collisional systems. 

The chosen molecular system (\ce{HCO+} and \ce{He}) provided an excellent test case for the study of long-range effects and 
led to a complete derivation of all the relevant scattering parameters at the best accuracy achieved so far.

The validation of the resulting data with the experimental counterparts and the computational results from previous works led to a further proof of the accuracy of the computational procedure. 
For instance, the comparison of the rotational frequencies of the bound state obtained by means of a pure computational methodology 
with the corresponding experimental values, revealed a percentage error always lower than 0.12\%.
Furthermore, the obtained pressure coefficients are in good agreement with the previously experimentally measured values over three rotational transitions. 
However, a more thorough analysis of our computational performance would be more significant if a greater number of experimental transitions were accessible. 

\section*{Supplementary material}
See supplementary material for the complete list of the computed de-excitation  rate coefficients from~5 to~100\,K\@.

\begin{acknowledgments}
This study was supported by University of Bologna (RFO funds). The SMART@SNS Laboratory (http://smart.sns.it) is
acknowledged for providing high-performance computing facilities. Support by the Italian Space Agency (ASI; ‘Life in Space’ project, N. 2019-3-U.0) is also acknowledged.

François Lique acknowledges financial support from the Institut Universitaire de France and the Programme National “Physique et Chimie du Milieu Interstellaire” (PCMI) of CNRS/INSU with INC/INP cofunded by CEA and CNES.
\end{acknowledgments}

\section*{Data Availability Statement}
The data that support the findings of this study are available within the article and its supplementary material.


\begin{thebibliography}{10}

\bibitem{roueff2013molecular}
E.~Roueff and F.~Lique, ``{Molecular excitation in the interstellar medium:
  Recent advances in collisional, radiative, and chemical processes},'' {\em
  Chemical reviews}, vol.~113, no.~12, pp.~8906--8938, 2013.

\bibitem{kauffmann2017molecular}
J.~Kauffmann, P.~F. Goldsmith, G.~Melnick, V.~Tolls, A.~Guzman, and K.~M.
  Menten, ``{Molecular Line Emission as a Tool for Galaxy Observations
  (LEGO)-I. HCN as a tracer of moderate gas densities in molecular clouds and
  galaxies},'' {\em Astronomy \& Astrophysics}, vol.~605, p.~L5, 2017.

\bibitem{adler2007simple}
T.~B. Adler, G.~Knizia, and H.-J. Werner, ``{A simple and efficient CCSD(T)-F12
  approximation},'' 2007.

\bibitem{knizia2009simplified}
G.~Knizia, T.~B. Adler, and H.-J. Werner, ``{Simplified CCSD(T)-F12 methods:
  Theory and benchmarks},'' {\em The Journal of Chemical Physics}, vol.~130,
  no.~5, p.~054104, 2009.

\bibitem{peterson2008systematically}
K.~A. Peterson, T.~B. Adler, and H.-J. Werner, ``{Systematically convergent
  basis sets for explicitly correlated wavefunctions: The atoms H, He, B--Ne,
  and Al--Ar},'' {\em The Journal of Chemical Physics}, vol.~128, no.~8,
  p.~084102, 2008.

\bibitem{kendall1992electron}
R.~A. Kendall, T.~H. Dunning~Jr, and R.~J. Harrison, ``{Electron affinities of
  the first-row atoms revisited. Systematic basis sets and wave functions},''
  {\em The Journal of Chemical Physics}, vol.~96, no.~9, pp.~6796--6806, 1992.

\bibitem{lique2010benchmarks}
F.~Lique, J.~K{\l}os, and M.~Hochlaf, ``{Benchmarks for the generation of
  interaction potentials for scattering calculations: applications to
  rotationally inelastic collisions of C$_4$ (X$^3$ $\Sigma^-_g$) with He},''
  {\em Physical Chemistry Chemical Physics}, vol.~12, no.~48, pp.~15672--15680,
  2010.

\bibitem{ajili2013accuracy}
Y.~Ajili, K.~Hammami, N.~E. Jaidane, M.~Lanza, Y.~N. Kalugina, F.~Lique, and
  M.~Hochlaf, ``{On the accuracy of explicitly correlated methods to generate
  potential energy surfaces for scattering calculations and clustering:
  application to the HCl--He complex},'' {\em Physical Chemistry Chemical
  Physics}, vol.~15, no.~25, pp.~10062--10070, 2013.

\bibitem{petrie2007ions}
S.~Petrie and D.~K. Bohme, ``{Ions in space},'' {\em Mass spectrometry
  reviews}, vol.~26, no.~2, pp.~258--280, 2007.

\bibitem{herbst2005molecular}
E.~Herbst, ``{Molecular ions in interstellar reaction networks},'' in {\em
  Journal of Physics: Conference Series}, vol.~4, p.~003, IOP Publishing, 2005.

\bibitem{snyder1976detection}
L.~Snyder, J.~Hollis, F.~Lovas, and B.~Ulich, ``{Detection, identification, and
  observations of interstellar HC$^{13}$O$^+$},'' {\em The Astrophysical
  Journal}, vol.~209, pp.~67--74, 1976.

\bibitem{langer1978observations}
W.~Langer, R.~Wilson, P.~Henry, and M.~Guelin, ``{Observations of anomalous
  intensities in the lines of the HCO$^+$ isotopes},'' {\em The Astrophysical
  Journal}, vol.~225, pp.~L139--L142, 1978.

\bibitem{welch1981millimeter}
W.~Welch, M.~Wright, R.~Plambeck, J.~Bieging, and B.~Baud,
  ``{Millimeter-wavelength aperture synthesis of molecular lines toward Orion
  KL},'' {\em The Astrophysical Journal}, vol.~245, pp.~L87--L90, 1981.

\bibitem{lattanzi2007rotational}
V.~Lattanzi, A.~Walters, B.~J. Drouin, and J.~C. Pearson, ``Rotational spectrum
  of the formyl cation, hco+, to 1.2 thz,'' {\em The Astrophysical Journal},
  vol.~662, no.~1, p.~771, 2007.

\bibitem{caselli1998ionization}
P.~Caselli, C.~Walmsley, R.~Terzieva, and E.~Herbst, ``{The ionization fraction
  in dense cloud cores},'' {\em The Astrophysical Journal}, vol.~499, no.~1,
  p.~234, 1998.

\bibitem{buhl1970unidentified}
D.~Buhl and L.~Snyder, ``{Unidentified interstellar microwave line},'' {\em
  Nature}, vol.~228, no.~5268, pp.~267--269, 1970.

\bibitem{article}
R.~Woods, T.~Dixon, R.~Saykally, and P.~Szanto, ``{Laboratory Microwave
  Spectrum of HCO$^{+}$},'' {\em Physical Review Letters}, vol.~35,
  pp.~1269--1272, 11 1975.

\bibitem{denis2020rotational}
O.~Denis-Alpizar, T.~Stoecklin, A.~Dutrey, and S.~Guilloteau, ``{Rotational
  relaxation of HCO+ and DCO+ by collision with H2},'' {\em Monthly Notices of
  the Royal Astronomical Society}, vol.~497, no.~4, pp.~4276--4281, 2020.

\bibitem{monteiro1985rotational}
T.~Monteiro, ``{Rotational excitation of HCO$^+$ by collisions with H2},'' {\em
  Monthly Notices of the Royal Astronomical Society}, vol.~214, no.~4,
  pp.~419--427, 1985.

\bibitem{buffa2008experimental}
G.~Buffa, L.~Dore, F.~Tinti, and M.~Meuwly, ``{Experimental and theoretical
  study of helium broadening and shift of HCO$^+$ rotational lines},'' {\em
  ChemPhysChem}, vol.~9, no.~15, pp.~2237--2244, 2008.

\bibitem{raghavachari1989fifth}
K.~Raghavachari, G.~W. Trucks, J.~A. Pople, and M.~Head-Gordon, ``{A
  fifth-order perturbation comparison of electron correlation theories},'' {\em
  Chemical Physics Letters}, vol.~157, no.~6, pp.~479--483, 1989.

\bibitem{peterson1994benchmark}
K.~A. Peterson, D.~E. Woon, and T.~H. Dunning~Jr, ``{Benchmark calculations
  with correlated molecular wave functions. IV. The classical barrier height of
  the H + H$_2$ -- H$_2$ + H reaction},'' {\em The Journal of Chemical
  Physics}, vol.~100, no.~10, pp.~7410--7415, 1994.

\bibitem{buffa2009state}
G.~Buffa, L.~Dore, and M.~Meuwly, ``{State-to-state rotational transition rates
  of the HCO$^+$ ion by collisions with helium},'' {\em Monthly Notices of the
  Royal Astronomical Society}, vol.~397, no.~4, pp.~1909--1914, 2009.

\bibitem{Salomon2019}
T.~Salomon, M.~Töpfer, P.~Schreier, S.~Schlemmer, H.~Kohguchi, L.~Surin, and
  O.~Asvany, ``{Double resonance rotational spectroscopy of He–HCO$^+$},''
  {\em Physical Chemistry Chemical Physics}, vol.~21, pp.~3440--3445, 2019.

\bibitem{davies1984diode}
P.~Davies and W.~Rothwell, ``{Diode laser detection of the bending mode of
  HCO$^+$},'' {\em The Journal of Chemical Physics}, vol.~81, no.~12,
  pp.~5239--5240, 1984.

\bibitem{dore2003struct}
L.~{Dore}, S.~{Beninati}, C.~{Puzzarini}, and G.~{Cazzoli}, ``"{Study of
  vibrational interactions in DCO$^{+}$ by millimeter-wave spectroscopy and
  determination of the equilibrium structure of the formyl ion}",'' {\em The
  Journal of Chemical Physics}, vol.~118, pp.~7857--7862, May 2003.

\bibitem{dunning1989gaussian}
T.~H. Dunning~Jr, ``{Gaussian basis sets for use in correlated molecular
  calculations. I. The atoms boron through neon and hydrogen},'' {\em The
  Journal of Chemical Physics}, vol.~90, no.~2, pp.~1007--1023, 1989.

\bibitem{woon1993gaussian}
D.~E. Woon and T.~H. Dunning~Jr, ``{Gaussian basis sets for use in correlated
  molecular calculations. III. The atoms aluminum through argon},'' {\em The
  Journal of Chemical Physics}, vol.~98, no.~2, pp.~1358--1371, 1993.

\bibitem{papajak2011perspectives}
E.~Papajak, J.~Zheng, X.~Xu, H.~R. Leverentz, and D.~G. Truhlar,
  ``{Perspectives on basis sets beautiful: seasonal plantings of diffuse basis
  functions},'' {\em Journal of Chemical Theory and Computation}, vol.~7,
  no.~10, pp.~3027--3034, 2011.

\bibitem{alessandrini2019extension}
S.~Alessandrini, V.~Barone, and C.~Puzzarini, ``{Extension of the “Cheap”
  Composite Approach to Noncovalent Interactions: The jun-ChS Scheme},'' {\em
  Journal of Chemical Theory and Computation}, vol.~16, no.~2, pp.~988--1006,
  2019.

\bibitem{watts1993coupled}
J.~D. Watts, J.~Gauss, and R.~J. Bartlett, ``{Coupled-cluster methods with
  noniterative triple excitations for restricted open-shell Hartree--Fock and
  other general single determinant reference functions. Energies and analytical
  gradients},'' {\em The Journal of Chemical Physics}, vol.~98, no.~11,
  pp.~8718--8733, 1993.

\bibitem{werner2012wires}
H.~J. Werner, P.~J. Knowles, G.~Knizia, F.~R. Manby, and M.~Schütz, ``{Molpro:
  a general-purpose quantum chemistry program package},'' {\em WIREs Comput.
  Mol. Sci.}, vol.~2, pp.~242--253, 2012.

\bibitem{feller1992application}
D.~Feller, ``{Application of systematic sequences of wave functions to the
  water dimer},'' {\em The Journal of Chemical Physics}, vol.~96, no.~8,
  pp.~6104--6114, 1992.

\bibitem{helgaker1997basis}
T.~Helgaker, W.~Klopper, H.~Koch, and J.~Noga, ``{Basis-set convergence of
  correlated calculations on water},'' {\em The Journal of Chemical Physics},
  vol.~106, no.~23, pp.~9639--9646, 1997.

\bibitem{boys1970calculation}
S.~F. Boys and F.~Bernardi, ``{The calculation of small molecular interactions
  by the differences of separate total energies. Some procedures with reduced
  errors},'' {\em Molecular Physics}, vol.~19, no.~4, pp.~553--566, 1970.

\bibitem{lique2019gas}
F.~Lique and A.~Faure, {\em {Gas-Phase Chemistry in Space; From elementary
  particles to complex organic molecules}}.
\newblock 2019.

\bibitem{hutson2019user}
J.~M. Hutson and C.~Sueur, ``{User Manual for MOLSCAT, BOUND and FIELD, Version
  2020.0: programs for quantum scattering properties and bound states of
  interacting pairs of atoms and molecules},'' {\em arXiv preprint
  arXiv:1903.06755}, 2019.

\bibitem{cazzoli2012hco+}
G.~{Cazzoli}, L.~{Cludi}, G.~{Buffa}, and C.~{Puzzarini}, ``"{Precise THz
  Measurements of HCO$^{+}$, N$_{2}$H$^{+}$, and CF$^{+}$ for Astrophysical
  Observations}",'' {\em Astroph. J. Supp.}, vol.~203, p.~11, Nov. 2012.

\bibitem{alexander1984hybrid}
M.~H. Alexander, ``{Hybrid quantum scattering algorithms for long-range
  potentials},'' {\em The Journal of Chemical Physics}, vol.~81, no.~10,
  pp.~4510--4516, 1984.

\bibitem{alexander1987stable}
M.~H. Alexander and D.~E. Manolopoulos, ``{A stable linear reference potential
  algorithm for solution of the quantum close-coupled equations in molecular
  scattering theory},'' {\em The Journal of Chemical Physics}, vol.~86, no.~4,
  pp.~2044--2050, 1987.

\bibitem{manolopoulos1986improved}
D.~Manolopoulos, ``{An improved log derivative method for inelastic
  scattering},'' {\em The Journal of Chemical Physics}, vol.~85, no.~11,
  pp.~6425--6429, 1986.

\bibitem{yazidi2014revised}
O.~Yazidi, D.~Ben~Abdallah, and F.~Lique, ``{Revised study of the collisional
  excitation of HCO$^+$ by H2 (j= 0)},'' {\em Monthly Notices of the Royal
  Astronomical Society}, vol.~441, no.~1, pp.~664--670, 2014.

\bibitem{woods1988microwave}
R.~Woods, ``{Microwave spectroscopy of molecular ions in the laboratory and in
  space},'' {\em Philosophical Transactions of the Royal Society of London.
  Series A, Mathematical and Physical Sciences}, vol.~324, no.~1578,
  pp.~141--146, 1988.

\bibitem{feller2000ccsdt}
D.~Feller and J.~A. Sordo, ``{A CCSDT study of the effects of higher order
  correlation on spectroscopic constants. I. First row diatomic hydrides},''
  {\em The Journal of Chemical Physics}, vol.~112, no.~13, pp.~5604--5610,
  2000.

\bibitem{blake1987laboratory}
G.~A. Blake, K.~Laughlin, R.~Cohen, K.~L. Busarow, and R.~Saykally,
  ``{Laboratory measurement of the pure rotational spectrum of vibrationally
  excited HCO$^+$ (v$_2= 1$) by far-infrared laser sideband spectroscopy},''
  {\em Astrophysical Journal Letters}, vol.~316, pp.~L45--L48, 1987.

\bibitem{yousaf2008optimized}
K.~E. Yousaf and K.~A. Peterson, ``{Optimized auxiliary basis sets for
  explicitly correlated methods},'' {\em The Journal of Chemical Physics},
  vol.~129, no.~18, p.~184108, 2008.

\bibitem{kraemer1976identification}
W.~Kraemer and G.~Diercksen, ``{Identification of interstellar X-ogen as
  HCO$^+$},'' {\em The Astrophysical Journal}, vol.~205, pp.~L97--L100, 1976.

\bibitem{dalgarno1973chemiionization}
A.~Dalgarno, M.~Oppenheimer, and R.~Berry, ``{Chemiionization in interstellar
  clouds},'' {\em The Astrophysical Journal}, vol.~183, p.~L21, 1973.

\bibitem{rimola2021interaction}
A.~Rimola, C.~Ceccarelli, N.~Balucani, and P.~Ugliengo, ``{Interaction of
  HCO$^+$ cations with interstellar negative grains. Quantum chemical
  investigation and astrophysical implications},'' {\em Frontiers in Astronomy
  and Space Sciences}, vol.~8, p.~38, 2021.

\end{thebibliography}

\end{document}


\preprint{AIP/123-QED}

\title[\ce{HCO+} and \ce{He} collisional system]{Seeking for accurate scattering parameters of the \ce{HCO+} and \ce{He} collisional system: a new efficient computational  potential energy surface.}
\author{F. Tonolo}
 \affiliation{Scuola Normale Superiore, Piazza dei Cavalieri 7, I-56126 Pisa.}
 \affiliation{Dipartimento di Chimica “Giacomo Ciamician”, Università di Bologna, Via F. Selmi 2, I-40126 Bologna, Italy}
 
\author{L. Bizzocchi}%
 \affiliation{Scuola Normale Superiore, Piazza dei Cavalieri 7, I-56126 Pisa.}
 \affiliation{Dipartimento di Chimica “Giacomo Ciamician”, Università di Bologna, Via F. Selmi 2, I-40126 Bologna, Italy}
 
\author{L. Dore}
 \affiliation{Dipartimento di Chimica “Giacomo Ciamician”, Università di Bologna, Via F. Selmi 2, I-40126 Bologna, Italy}
 
\author{F. Lique}
 \affiliation{Univ. Rennes, CNRS, IPR (Institut de Physique de Rennes) – UMR 6251, F-35000 Rennes, France}
\author{C. Puzzarini}
 \affiliation{Dipartimento di Chimica “Giacomo Ciamician”, Università di Bologna, Via F. Selmi 2, I-40126 Bologna, Italy}
 
\author{V. Barone}
 \affiliation{Scuola Normale Superiore, Piazza dei Cavalieri 7, I-56126 Pisa.}

 \homepage{http://www.Second.institution.edu/~Charlie.Author.}

\date{\today}
             
\maketitle    

\section*{Supplementary material}

\begin{table*}
\renewcommand{\tabcolsep}{3pt}
\renewcommand{\arraystretch}{1.1}
    \scriptsize
    \caption{Integrated rate coefficients $j \rightarrow j^{\prime}$ from 5 to 50 K. Units are $10^{-10}$ cm$^3$ s$^{-1}$.}
    \footnotesize
    \centering 
    \begin{threeparttable} 
    \begin{tabular}{ldddddddddd}
        \toprule
        \multirow{2}{*}{$j \rightarrow j^{\prime}$} & \multicolumn{10}{c}{Temperature / $K$} \\
        \cline{2-11}
        
         & \multicolumn{1}{c}{5 } & \multicolumn{1}{c}{10} & \multicolumn{1}{c}{15} & \multicolumn{1}{c}{20} & \multicolumn{1}{c}{25} & \multicolumn{1}{c}{30} & \multicolumn{1}{c}{35} & \multicolumn{1}{c}{40} & \multicolumn{1}{c}{45} & \multicolumn{1}{c}{50} \\
        \midrule
        1 $\rightarrow$ 0   & 1.368 & 1.188 & 1.104 &	1.056 &	1.026 &	1.006 &	0.992 &	0.983 &	0.977 &	0.972  \\                                    
        2 $\rightarrow$ 0   & 0.687 & 0.625 & 0.577 &	0.542 &	0.516 &	0.495 &	0.477 &	0.462 &	0.448 &	0.436  \\                                    
        2 $\rightarrow$ 1   & 1.670 & 1.641 & 1.594 &	1.556 &	1.527 &	1.504 &	1.487 &	1.473 &	1.463 &	1.456  \\                                    
        3 $\rightarrow$ 0   & 0.278 & 0.277 & 0.266 &	0.254 &	0.243 &	0.234 &	0.226 &	0.220 &	0.215 &	0.210  \\                                    
        3 $\rightarrow$ 1   & 1.030 & 1.002 & 0.966 &	0.929 &	0.895 &	0.865 &	0.838 &	0.814 &	0.792 &	0.773  \\                                    
        3 $\rightarrow$ 2   & 1.683 & 1.659 & 1.635 &	1.618 &	1.606 &	1.598 &	1.593 &	1.591 &	1.592 &	1.594  \\                                    
        4 $\rightarrow$ 0   & 0.185 & 0.177 & 0.169 &	0.162 &	0.155 &	0.150 &	0.145 &	0.140 &	0.137 &	0.133  \\                                    
        4 $\rightarrow$ 1   & 0.466 & 0.475 & 0.479 &	0.479 &	0.477 &	0.475 &	0.474 &	0.472 &	0.471 &	0.471  \\                                    
        4 $\rightarrow$ 2   & 1.177 & 1.161 & 1.128 &	1.095 &	1.065 &	1.039 &	1.015 &	0.993 &	0.973 &	0.955  \\                                    
        4 $\rightarrow$ 3   & 1.711 & 1.680 & 1.648 &	1.628 &	1.616 &	1.610 &	1.609 &	1.609 &	1.612 &	1.617  \\
        5 $\rightarrow$ 0   & 0.146 & 0.139 & 0.138 &	0.138 &	0.139 &	0.140 &	0.141 &	0.141 &	0.142 &	0.142  \\
        5 $\rightarrow$ 1   & 0.412 & 0.388 & 0.373 &	0.364 &	0.356 &	0.350 &	0.344 &	0.338 &	0.333 &	0.328  \\
        5 $\rightarrow$ 2   & 0.675 & 0.665 & 0.660 &	0.658 &	0.657 &	0.656 &	0.654 &	0.653 &	0.652 &	0.650  \\
        5 $\rightarrow$ 3   & 1.122 & 1.143 & 1.139 &	1.125 &	1.109 &	1.091 &	1.073 &	1.055 &	1.039 &	1.023  \\
        5 $\rightarrow$ 4   & 1.767 & 1.714 & 1.653 &	1.608 &	1.579 &	1.561 &	1.550 &	1.546 &	1.545 &	1.547  \\
        \bottomrule
    \end{tabular}

    \end{threeparttable}
    \label{rates_all}
    \end{table*}

\begin{table*}
\renewcommand{\tabcolsep}{3pt}
\renewcommand{\arraystretch}{1.1}
    \scriptsize
    \caption{Integrated rate coefficients $j \rightarrow j^{\prime}$ from 60 to 100 K. Units are $10^{-10}$ cm$^3$ s$^{-1}$.}
    \footnotesize
    \centering 
    \begin{threeparttable} 
    \begin{tabular}{lddddd}
        \toprule
        \multirow{2}{*}{$j \rightarrow j^{\prime}$} & \multicolumn{5}{c}{Temperature / $K$} \\
        \cline{2-6}
        
         & \multicolumn{1}{c}{60 } & \multicolumn{1}{c}{70} & \multicolumn{1}{c}{80} & \multicolumn{1}{c}{90} & \multicolumn{1}{c}{100} \\
        \midrule
        1 $\rightarrow$ 0   & 0.968 &	0.967 &	0.967 &	0.968 &	0.969  \\                                    
        2 $\rightarrow$ 0   & 0.416 &	0.398 &	0.384 &	0.372 &	0.362  \\                                    
        2 $\rightarrow$ 1   & 1.446 &	1.443 &	1.443 &	1.446 &	1.450  \\                                    
        3 $\rightarrow$ 0   & 0.204 &	0.200 &	0.198 &	0.197 &	0.197  \\                                    
        3 $\rightarrow$ 1   & 0.740 &	0.714 &	0.693 &	0.676 &	0.661  \\                                    
        3 $\rightarrow$ 2   & 1.601 &	1.612 &	1.624 &	1.637 &	1.650  \\                                    
        4 $\rightarrow$ 0   & 0.128 &	0.124 &	0.122 &	0.120 &	0.118  \\                                    
        4 $\rightarrow$ 1   & 0.471 &	0.473 &	0.476 &	0.478 &	0.481  \\                                    
        4 $\rightarrow$ 2   & 0.924 &	0.898 &	0.877 &	0.858 &	0.842  \\                                    
        4 $\rightarrow$ 3   & 1.628 &	1.642 &	1.657 &	1.672 &	1.686  \\
        5 $\rightarrow$ 0   & 0.143 &	0.143 &	0.143 &	0.143 &	0.143  \\
        5 $\rightarrow$ 1   & 0.320 &	0.313 &	0.307 &	0.302 &	0.297  \\
        5 $\rightarrow$ 2   & 0.648 &	0.647 &	0.646 &	0.645 &	0.644  \\
        5 $\rightarrow$ 3   & 0.994 &	0.969 &	0.948 &	0.929 &	0.913  \\
        5 $\rightarrow$ 4   & 1.556 &	1.569 &	1.585 &	1.601 &	1.617  \\
        \bottomrule
    \end{tabular}

    \end{threeparttable}
    \label{rates_all}
    \end{table*}

\clearpage
\bibliography{aipsamp}